\documentclass[aps,prl,twocolumn,10pt,superscriptaddress,showpacs]{revtex4-1}
\usepackage{amsmath,amssymb,bm,graphicx,color}
\usepackage{comment}

\makeatletter
\AtBeginDocument{\@ifpackageloaded{natbib}{\ifNAT@numbers\if@filesw\immediate\write\@auxout{\string\global\string\NAT@numberstrue}\fi\fi}{}}
\makeatother
\begin{document}

\title{Anomalous Josephson current in 
 superconducting topological insulators}

\author{Ai Yamakage}
\affiliation{Department of Applied Physics, Nagoya University, Nagoya 464-8603, Japan}

\author{Masatoshi Sato}
\affiliation{Department of Applied Physics, Nagoya University, Nagoya 464-8603, Japan}

\author{Keiji Yada}
\affiliation{Department of Applied Physics, Nagoya University, Nagoya 464-8603, Japan}

\author{Satoshi Kashiwaya}
\affiliation{National Institute of Advanced Industrial Science and Technology (AIST), Tsukuba 305-8568, Japan}

\author{Yukio Tanaka}
\affiliation{Department of Applied Physics, Nagoya University, Nagoya 464-8603, Japan}
\date{\today}

\begin{abstract}
We investigate the effect of helical Majorana fermions
at the surface of superconducting topological insulators (STI)
on the Josephson current by referring to
possible pairing states of Cu-doped Bi$_2$Se$_3$. 
The surface state in the present STI has a spin helicity 
because the directions of spin and momentum are locked to each other.
The Josephson current-phase relation 
in an STI/$s$-wave superconductor junction shows robust $\sin(2\varphi)$ owing
to mirror symmetry, where 
$\varphi$ denotes the macroscopic phase difference 
between the two superconductors. 
In contrast, the maximum Josephson current in an STI/STI junction
exhibits a nonmonotonic temperature dependence
depending on the relative spin helicity of the two surface states.
Detecting these features qualifies as  distinct experimental evidence 
for the identification of the helical Majorana fermion in STIs. 
%Majorana nature of the Andreev bound states.
\end{abstract}

%74.45.+c Proximity effect; Andreev reflection; SN and SNS junctions  
%73.20.At Surface states, band structure, electron density of states  
%03.65.Vf  Topological phases(quantum mechanics)

\maketitle

{\it Introduction.}
The Josephson effect is one of the most important 
quantum phenomena in superconductivity.  
As a phase-sensitive probe of the superconducting state, 
the Josephson effect has contributed to the identification of unconventional superconductivity: 
It is well known that a $\pi$ phase shift \cite{golubov04} can be used for Josephson junction interferometers \cite{harlingen95,Tsuei}, 
which established the  $d$-wave symmetry of cuprates \cite{harlingen95,Tsuei,Sigrist95}.
Furthermore, there have been several theoretical 
studies on the anomalous features of 
spin-triplet superconductor junctions 
\cite{pals76,pals77,geshkenbein86,yip93,tanaka94,tanaka96,Josephson3,tanaka97}. 
In particular, an unusual current-phase relation 
\cite{pals76,pals77,geshkenbein86,yip93,tanaka94,tanaka96,Fogelstrom,Radovic1,
Radovic2,Radovic3,Radovic4} 
or 
a nonmonotonic temperature dependence of the maximum Josephson current 
\cite{tanaka96,Josephson3,tanaka97} 
leads to the manifestation of unconventional pairing states 
\cite{golubov04,Kashiwaya00,Hu,TK95,Kawabata05}. \par

Recently, a new superconductor Cu-doped Bi$_2$Se$_3$ has been discovered \cite{hor10}. 
Since
the undoped material, Bi$_2$Se$_3$, 
is a topological insulator with gapless surface Dirac fermions \cite{hasan10,qi11,hasan11},
this superconductor is dubbed as a superconducting topological insulator (STI).
The STI is one of the candidates for topological superconductors, which has Majorana fermions as gapless surface Andreev bound states (ABSs) 
\cite{wilczek09,qi11,tanaka12,alicea12,Sato09,Sato10}. 
Indeed,
the presence of a zero-bias conductance peak \cite{TK95} in point-contact tunneling spectroscopy experiments has suggested the topological superconductivity in this material \cite{sasaki11,koren11,kirzhner12,koren12,peng13}. 
Recent scanning tunneling spectroscopy experiments, however, show a conflicting result indicating nontopological $s$-wave superconductivity \cite{Levy}.
Also, the appearance of a zero-bias conductance peak could be explained by nontopological origins \cite{BTK,Deutscher,Kastalsky,Beenakker}.

The Josephson effect resolves this controversy: 
Experiments of current-phase relation of the Josephson junction enables us to determine the parity of the pair potential.
If the parity of the pairing potential is odd (even),
then the topological (nontopological) phase is realized \cite{Sato09,Sato10,fu10}.
In addition, 
the temperature dependence of the Josephson current at low temperatures 
can verify
the existence of gapless ABSs  in the topological phase
in a manner similar to $d$-wave superconductor junctions \cite{Hu, TK95}.
%
%
%
%
% The latter has an ABS called a helical Majorana fermion in which the directions of spin and momentum are locked to each other by the strong spin-orbit coupling without breaking time-reversal symmetry. 
% In addition, ABSs, which appear in the topological phase in STIs, can be hybridized with surface Dirac fermions originating from the parent topological insulator Bi$_{2}$Se$_{3}$ \cite{hao11,hsieh12,yamakage12}. This specific feature of the energy dispersion can enhance the surface density of state (SDOS) at the Fermi energy, which is expected to affect the Josephson current through the surface \cite{hsieh12,yamakage12}. 
% For high-$T_{c}$ cuprates, ABSs \cite{Kashiwaya00,Hu,TK95} with flat band dispersion induce several interesting Josephson current features \cite{golubov04,tanaka96,Josephson3,tanaka97,Kawabata05}. 
% However, there has been no microscopic calculation of the Josephson current via the helical Majorana fermion expected in an STI. \par
% Thus, predicting the Josephson current of STI is essential for establishing its topological superconductivity. 

In this Rapid Communication, we calculate the Josephson current 
of STIs for both topological and nontopological phases 
based on a microscopic Hamiltonian, taking into account the 
surface Dirac fermion specific to them. 
The properties of the Josephson current $J$ in 
the nontopological phase are 
conventional, and the current-phase ($\varphi)$ relation becomes 
$\sin \varphi$.  
However, in the topological phase,
we clarify that the first-order component, $\sin\varphi$, vanishes.
% even in the presence of spin-orbit scattering at the interface
% owing to the antisymmetric property of the mirror plane.
Then, $J$ between an $s$-wave superconductor and the STI
exhibits robust second-order behavior: $J \propto \sin 2\varphi$.
Furthermore, we find that 
gapless ABSs in the topological phase
cause a distinct temperature dependence of the Josephson current between two STIs.
Detection of these  features will establish the
topological superconductivity in STIs.

{\it Microscopic calculation.}
In the following, we calculate the Josephson current in an STI microscopically,
using the three-dimensional lattice model that takes into
account proper STI electric structures. 
%since we must take into account of the 
%specific electronic structures in the topological insulator.  
For the STI pairing symmetry \cite{fu10}, 
there are four possible gap functions derived from 
the crystal symmetry.  
Among them, we focus on the two kinds of full gap functions, $\Delta_1$
and $\Delta_2$, 
that belong to the $A_{1g}$
and $A_{1u}$ representations in the $D_{3d}$ point group, respectively.
These pairing symmetries are consistent with  
specific heat measurements \cite{Kriener}.  
Both of them are invariant under time reversal, but the parities under the
inversion are different:
%and the gap
%function is odd under the mirror reflection.
whereas the gap function $\Delta_1$ is an 
intra-orbital pairing and has  
even parity under the inversion, $\Delta_2$ is 
an intra-orbital pairing and has  
odd parity under the inversion.  
From these differences, 
only the latter ($\Delta_2$) supports the 
topological superconductivity accompanying Majorana fermions on its
surface \cite{Sato09, Sato10, fu10}.

An STI has a diagonal mirror plane $(x,y,z) \to (-x,y,z)$ in its
crystal structure.  
The gap function $\Delta_1$ ($\Delta_2$) is even (odd) under mirror
reflection.
This mirror symmetry is also important for its topological properties.
For instance, the nontrivial mirror symmetry of $\Delta_2$ yields a structural
transition in the energy dispersion of 
Majorana fermions, which enables us to explain the zero-bias
peak of the tunneling conductance observed experimentally \cite{yamakage12}.
Below, we shall argue how the mirror symmetry affects the
Josephson current.
%We denote the gap function for the former and the latter 
%representation as 
%$\Delta_{ntp}$ and $\Delta_{tp}$, respectively. 
% 
%we focus on the full-gap intra-orbital and inter-orbital 
%pairing without momentum dependence. 
%The irreducible representations of them are 
% classified as $A_{1g}$ and $A_{1u}$, 
%respectively. 
% representation which is a full-gap
%topological superconductor, while
%qualitatively similar results are obtained
%for the nodal topological superconductor in the $E_u$ representation.

\begin{figure}
\centering
\includegraphics[scale=1]{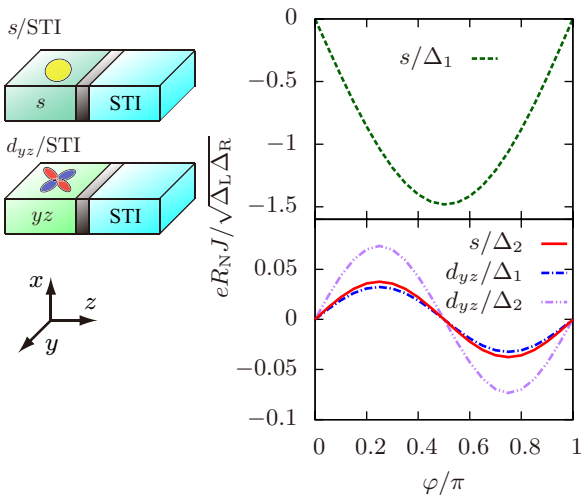}
\caption{
Geometries of STI junctions (left) and the corresponding current-phase relations (right).
$s$/STI junctions in the nontopological ($s/\Delta_1$) and topological ($s/\Delta_2$) phases and
$d_{yz}$/STI junctions in the nontopological ($d_{yz}/\Delta_1$) and topological ($d_{yz}/\Delta_2$) phases
 are assumed. 
We assume thin normal layers (black shaded region) at the center of 
the junctions.
The Josephson current flows along the $z$ direction.
We use the following parameters taken from Ref. \cite{hao11}:
$\mu = -0.5$, $t_x=0.1t_z$, $t_y=t_z$,  $t = 0.1 t_z$,
$\Delta_{\rm STI} = 0.1$, $\Delta_{s} = \Delta_d = 0.2$, $m_0 = -0.7$, $m_1=0.5$, $m_2 = 1.5$, $\mu_{\rm STI} = 0.9$, $v_z = 1$, $v = 1.5$, $N_{\rm C} = 2$, and $T=0$. $N_{\rm L}$ and $N_{\rm R}$ take sufficiently large values ($\sim \negthickspace40 \text{--} 80$) to converge the calculation.
}
\label{s-wave}
\end{figure}

\begin{figure}
\centering
\includegraphics{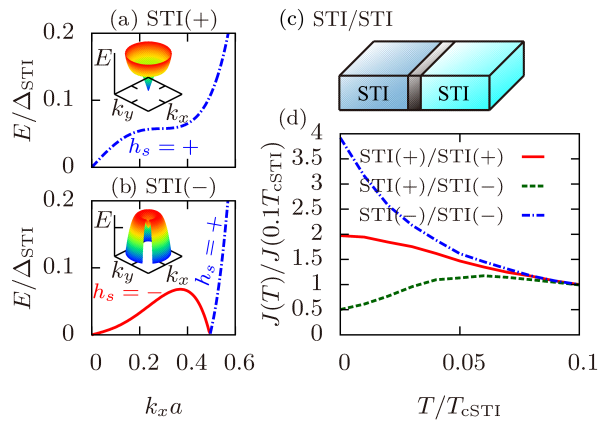}
\caption{
Energy spectra and temperature dependencies of Josephson current $J(T)$.
(a)
The positive energy part of the surface state for $\mu_{\rm STI}=v_z$ in the region of $k_y=0$ and $k_x>0$.
The inset shows the corresponding overall spectrum of the $h_s=+$ branch of the surface state.
(b)
The positive energy part of the surface
  state for $\mu_{\rm STI}=0.9v_z$ in the region of $k_y=0$ and $k_x>0$.
The inset shows the corresponding overall spectrum of the $h_s=-$ branch of
 the surface state, 
which is discussed in Ref. \cite{yamakage12}.
(c) 
Geometry of an STI/STI junction.
(d)
Temperature dependencies of Josephson current $J(T)$ at STI/STI junctions.
Here, the STI(+)/STI(+) [STI($-$)/STI($-$)] junction is prepared by choosing the
 chemical potential as $\mu_{\rm STI}^{\rm L}=\mu_{\rm STI}^{\rm R}=v_z$
 ($\mu_{\rm STI}^{\mathrm L} = \mu_{\rm STI}^{\rm R} = 0.9v_z$), where $\mu_{\rm
 STI}^{\mathrm L}$ and  $\mu_{\rm STI}^{\mathrm R}$ are the chemical
 potentials in the left and right STIs, respectively. 
The spin-helicity-mismatched case of the STI(+)/STI($-$) junction is constructed by
 setting $\mu_{\rm STI}^{\rm L} = v_z$ and $\mu_{\rm STI}^{\rm R}=0.9v_z$.
The values of hopping between N and the STI are chosen as 
 $t=0.1t_x$.
}
\label{Tdep}
\end{figure}

We consider various kinds of junctions, consisting of (a) an $s$-wave superconductor,   normal metal (N), and
an STI ($s$/STI); (b) a $d_{yz}$-wave superconductor, N, and an STI
($d_{yz}$/STI); 
and (c) an STI, N, and an STI (STI/STI). 
For (a) and (b), we consider both topological ($\Delta_2$) and nontopological ($\Delta_1$)
phases in the STI. 
In contrast, for case (c), 
we only consider the topological phase. 
The orientations of the superconductors at the junctions are chosen as
illustrated in Figs. \ref{s-wave} and \ref{Tdep}.

We use the following model in a cubic lattice in the calculation:
we put $s$, $d_{yz}$, or STI in the
left region ($1<z<N_{\rm L}$), N in the center region
($N_{\rm L} + 1 < z < N_{\rm L} + N_{\rm C}$), and  STI in the right
region ($N_{\rm L} + N_{\rm C} + 1 < z < N_{\rm L}+N_{\rm C}+N_{\rm
R}$).
Performing a Fourier transformation in the $x$ and $y$ directions effectively reduces the
Hamiltonian in each region  to a one-dimensional
lattice model in the $z$ direction,
${\cal H}({\bm k}_{\parallel})=\sum_{n_z,n_z'}
c^{\dagger}_{n_z, {\bm k}_{\parallel}}h_{n_z,n_z'}({\bm k}_{\parallel})
c_{n_z',{\bm k}_{\parallel}}
+\frac{1}{2}\sum_{n_z,n_z'}(
c^{\dagger}_{n_z, {\bm k}_{\parallel}}\Delta_{n_z,n_z'}({\bm k}_{\parallel})
c^{\dagger}_{n_z',-{\bm k}_{\parallel}}+{\rm H.c.}),
$
where ${\bm k}_{\parallel}=(k_x,k_y)$ is the momentum in the $x$ and
$y$ directions, and $c_{n_z, \bm k_\parallel}$ is the annihilation operator of an electron at site $n_z$ with $\bm k_\parallel$.
Here, the spin and orbital indices of the electron are implicit.
For $s$ and $d_{yz}$, $h_{n_z,n_z'}({\bm k}_{\parallel})$ and
$\Delta_{n_z,n_z'}({\bm k}_{\parallel})=i\psi_{n_z,n_z'}({\bm k}_{\parallel})s_y$ are  given by
\begin{eqnarray}
h_{n_z,n_z'}
&=& (2t_x \cos k_x + 2t_y \cos k_y - \mu)\delta_{n_z,n_z'}s_0
\label{eq:hsd}
\nonumber\\
&&+t_z(\delta_{n_z,n'_z+1} +\delta_{n_z+1,n'_z})s_0,
\\
\psi_{n_z,n_z'}&=&
\left\{
\begin{array}{lc}
\Delta_s \delta_{n_z,n_z'}, & \mbox{for $s,$ }\\
\Delta_d \sin k_y (\delta_{n_z+1,n_z'}-\delta_{n_z,n_z'+1})/i,
 & \mbox{for $d,$} 
\end{array}
\right. 
\end{eqnarray}
where $s_{\mu}=({\bm 1},{\bm s})$ is the Pauli matrix in  spin space.  
For N, 
$h_{n_z,n_z'}$ has the same form as Eq. (\ref{eq:hsd}), but
the pairing potential $\Delta_{n_z,n_z'}$ is zero.
%
%For simplicity, 
%we choose the parameters of $h_{n_z, n_z'}$ in N as the same value of
%$s$ ($d_{yz}$) for $s$/STI ($d_{yz}$/STI) junction.
For an STI, the electron has  additional orbital degrees of freedom
$\sigma=1,2$, and $h_{n_z,n_z'}$ and $\Delta_{n_z,n_z'}$ are given by
\begin{eqnarray}
h_{n_z,n_z'} &=& 
\{[m_0 + 2m_1 + 4m_2 -2m_2(\cos k_x + \cos k_y)] \sigma_x
\nonumber\\
&&+ v \sigma_z (s_y \sin k_x - s_x \sin k_y) - \mu_{\rm STI} \}\delta_{n_z,n_z'}s_0
\nonumber\\
&&
- m_1 \sigma_x (\delta_{n_z+1,n'_z} + \delta_{n_z,n'_z+1})s_0
\nonumber\\
&&
- i v_z \sigma_y/2 (\delta_{n_z+1,n'_z} - \delta_{n_z,n'_z+1})s_0,
\nonumber\\
\Delta_{n_z,n_z'}&=&
\begin{cases}
i\Delta_{\rm STI}\sigma_0 s_y \delta_{n_z, n_z'} \equiv \Delta_1, 
& \mbox{for nontop,} 
\\
i\Delta_{\rm STI}\sigma_y s_z s_y \delta_{n_z, n_z'} \equiv \Delta_2, 
& \mbox{for top,} 
\end{cases}
\end{eqnarray}
where $\sigma_{\mu}=({\bm 1}, {\bm \sigma})$ is the Pauli matrix in 
orbital space.
Note that $\Delta_2$ hosts topological superconductivity, but
$\Delta_1$ does not.
We assume that N is smoothly connected
to $s$ or $d_{yz}$ at the interface between them, and STI and N are
connected as
\begin{eqnarray}
 H_{{\rm N}\rightarrow{\rm STI}} 
&=& \sum_{\sigma s} t c^{\dag}_{n_z, {\bm k}_{\parallel}, s}
 c_{n_z+1,{\bm k}_{\parallel}, s, \sigma}+ \mathrm{H.c.},
\nonumber\\
 H_{{\rm STI}\rightarrow{\rm N}} 
&=& \sum_{\sigma s} t c^{\dag}_{n_z, {\bm k}_{\parallel}, s, \sigma}
 c_{n_z+1,{\bm k}_{\parallel}, s,}+ \mathrm{H.c.},
\end{eqnarray}
where $c_{n_z, {\bm k}_{\parallel}, s}$ 
($c_{n_z, {\bm k}_{\parallel}, s, \sigma}$) is the annihilation operator of
the electron in N (STI), and $s=\uparrow, \downarrow$ and $\sigma=1,2$ are
the spin and orbital indices.
%These do not depend on the spin due to the time-reversal symmetry.
%Pair potentials of $s$ ($\hat \Delta_s$), $d_{yz}$ ($\hat \Delta_{yz}$),
%gand STI ($\hat \Delta_{\rm STI}$) are given by 
%$
%\hat \Delta_s = \sum_{n=1}^{N_{\rm L}} c^\dag_{n} \Delta_s \tau_x c_n$,
%$\hat \Delta_{yz} = \sum_{n=1}^{N_{\rm L}-1}  c^\dag_n
% \frac{-i}{2}\Delta_{d} \sin k_y \tau_x c_{n+1} + \mathrm{H.c.},
%$
With these settings, the Josephson current density is calculated 
as
\begin{equation}
 J = \frac{i}{2N_\parallel^2} \sum_{\bm k_\parallel,s} \langle
 c^\dag_{n_z, \bm k_\parallel,s} t_z c_{n_z+1, \bm k_\parallel,s} \rangle
 + \mathrm{c.c.}, 
\end{equation}
where $N_\parallel^2$ is the number of unit cells in the $xy$ plane,
$n_z$ is a site in N, and $\langle \cdots \rangle$ indicates the thermal average.
We adopt  
$\Delta_{A=s,d,{\rm STI}} = \Delta_0 \tanh(1.74 \sqrt{T_{\rm c}/T-1})$ 
with $\Delta_0 = 1.76 T_{\rm c}$ as the temperature dependence of
the pair potentials, 
which can be justified in the weak-coupling limit \cite{muhlschlegel59,yokoyama12}.

The current-phase relations obtained for  
$s$/STI and 
$d_{yz}$/STI junctions are summarized in 
Fig. \ref{s-wave}.
Here, the vertical axis denotes $e R_{\rm N} J / \sqrt{\Delta_{\rm L}
\Delta_{\rm R}}$, where $R_{\rm N}$ is the zero-bias resistivity in the
normal state and
%($\Delta_s=\Delta_d=\Delta_{\rm STI}=0$)
$\Delta_{\rm L}$ 
and
$\Delta_{\rm R}$ are the magnitudes of the pairing functions in the left
and right sides, respectively; i.e., $\Delta_{\rm L} = \Delta_s$ for $s/$STI
junctions, $\Delta_{\rm L} = \Delta_d$ for $d_{yz}$/STI junctions, and
$\Delta_{\rm R} = \Delta_{\rm STI}$. 

First, we look at the case of $s/$STI junctions. 
As shown in Fig. \ref{s-wave}, 
when the STI is in the nontopological phase ($s/\Delta_1$), 
the corresponding Josephson current shows a conventional sinusoidal
dependence $J(\varphi)\sim \sin \varphi$, 
where $\varphi$ is a macroscopic phase of 
$\Delta_{\rm STI}=|\Delta_{\rm STI}| e^{i \varphi}$.
However, when STI is in the topological phase ($s/\Delta_2$),
the resulting Josephson current exhibits second-order behavior.
Therefore, the Josephson current behaves significantly differently in the
topological and nontopological phases.

For $d_{yz}$/STI junctions,
both the nontopological ($d_{yz}/\Delta_1$)
and topological ($d_{yz}/\Delta_2$) junctions
exhibit second-order Josephson current behavior: 
$J(\varphi)\sim \sin 2\varphi$. See Fig. \ref{s-wave}.
The magnitude of the Josephson currents at these junctions is enhanced
as the temperature decreases, 
owing to the existence
of the flat band ABS of $d_{yz}$-wave superconductivity \cite{Hu} at the
interface of the $d_{yz}$/STI junctions \cite{Kashiwaya00,tanaka94,TK95,tanaka96}.
%
%The magnitude of Josephson currents in these junctions are much smaller than
%that of $s/\Delta_1$ junction.
% but with the decrease of temperature. 
It is remarkable that the magnitude of the Josephson current at the
$d_{yz}/\Delta_2$ junction is larger than that at the $d_{yz}/\Delta_1$ junction.  
This is because there exists a helical ABS owing to the topological
superconductivity of $\Delta_2$, in addition to  
the flat band ABS at the interface of the $d_{yz}$-wave superconductor.
%
%consistent with junctions in high $T_c$ cuprate \cite{
%tanaka96}.  

{\it Symmetry-based argument.}
In the above, we found that the topological junctions, $s/\Delta_2$ and
$d_{yz}/\Delta_2$, exhibit robust $\sin 2\varphi$ Josephson
current behavior.
To understand this, 
we present here a general argument based on the symmetry of the system.
First, the Josephson current is generally decomposed into a series of
different orders \cite{TK95}:
\begin{equation}
J(\varphi)=\sum_{n=1}\left(J_n\sin n \varphi+I_n \cos n\varphi\right),
\end{equation}
where $J_n$ and $I_n$ decrease as $n$ increases.
Under time reversal, $J(\varphi)$ goes to $-J(-\varphi)$; thus
in time-reversal-symmetric junctions, $J(\varphi)$ satisfies
$
J(\varphi)=-J(-\varphi).
$
This implies that $I_n=0$ and the leading term is $J(\varphi)\sim
\sin \varphi$. 

If we take into account  mirror symmetry, however,
an additional constraint is required for $J(\varphi)$ \cite{yip90}.
% '±'±'ÉYipŽ'̘_•¶ˆø—p
At the $s/\Delta_2$ and $d_{yz}/\Delta_2$ junctions, 
%Consider a junction consisting of an $s$-wave superconductor and
%STI ($s$/STI) in the topological phase.
%Similar discussion is possible for both 
%$d_{yz}$/STI and $d_{y^{2}-z^{2}}$/STI junctions. 
the interface of the junctions is prepared so that the mirror plane of the STI
is perpendicular to it. 
Under mirror reflection, $x\rightarrow -x$,
the gap function $\Delta_2$ changes sign, whereas the $s$-wave and
$d_{yz}$ superconductors
do not.
Consequently, one obtains
an additional phase of $\pi$ in the Josephson current,
$J(\varphi+\pi)$, under mirror reflection.
Therefore, mirror symmetry implies 
\begin{eqnarray}
J(\varphi)=J(\varphi+\pi). 
\label{eq:jmirror}
\end{eqnarray}
This equation yields $J_{2n+1}=I_{2n+1}=0$ and the first-order term
 $J(\varphi)\sim \sin\varphi$ vanishes.
Consequently, the leading term becomes the second-order term, $J(\varphi)\sim \sin 2\varphi,$
 at these junctions, which reproduces our results qualitatively.
Here, note that the $\pi$ periodicity of the Josephson current in
 Eq. (\ref{eq:jmirror}) is consistent with the effective Josephson coupling
 $\propto [(\Delta_s^{*})^2\Delta_2^2+{\rm H.c.}]$ discussed in Ref. \cite{fu10}.

In a similar manner, the second-order behavior of the $d_{yz}/\Delta_1$
junction can be explained by another mirror reflection of $y\to -y$.
Under this mirror reflection, the $d_{yz}$ gap function reverses its sign, whereas
$\Delta_1$ does not.
Because this sign reversal gives rise to an additional phase of $\pi$ 
and yields Eq. (\ref{eq:jmirror}) again, the second-order behavior is obtained.

Our argument above implies that the second-order behavior obtained here
is robust as long as 
time-reversal and mirror symmetries are preserved.
In particular, it is not lost even when spin-orbit scattering is
present at the junction.
This is completely different from the second-order behaviors observed for chiral
$p$-wave superconductors.
It has been known that chiral $p$-wave superconductors such as 
Sr$_{2}$RuO$_{4}$ are topological superconductors supporting chiral
Majorana edge states \cite{Maeno2,Kashiwaya11}, and 
the current-phase relation between an $s$-wave superconductor and a
chiral $p$-wave superconductor is proportional to $\sin2\varphi$, if one
neglects the spin-orbit interaction \cite{Yamashiro,asano03}.
However, because the chiral $p$-wave superconductor is not odd under 
mirror reflection and it breaks time-reversal invariance as well, 
the second-order behavior is fragile. 
Actually, the
first-order term $\cos \varphi$ appears
immediately if one takes into account the spin-orbit interaction
\cite{asano03}. 

Because the mirror symmetry responsible for the second-order behavior of
the topological junction $s/\Delta_2$ and $d_{yz}/\Delta_2$ is different
from that of the nontopological junction $d_{yz}/\Delta_1$, one can easily
distinguish them by breaking the mirror symmetries in different
manners. Indeed, if one breaks the mirror symmetry of $x\to -x$
($y\rightarrow -y$) by applying magnetic fields in the $y$ ($x$) direction, the
second-order behavior of the topological (nontopological) junction
becomes obscure, whereas that of the nontopological (topological) one is not.

{\it Spin-helicity-dependent Josephson effect.}
Now, we clarify the Josephson 
effect intrinsic to the helical Majorana fermions of the STI in the topological phase.
As mentioned above, the superconducting state $\Delta_2$ supports
helical Majorana fermions on its surface. 
The effective Hamiltonian of the surface helical Majorana
fermion at the interface of the junctions (the $xy$ plane) is represented as
$
H_{\rm surf}({\bm k}_{\parallel})=v_{\rm surf}(k_xs_y-k_ys_x)
$
near ${\bm k}_{\parallel}={\bm 0}$, which leads to a linear dispersion
of the Majorana cone, $E_{\rm
surf}=\pm v_{\rm surf}k_{\parallel}$
with $k_{\parallel}=|{\bm k}_{\parallel}|$.
The spin and the momentum are locked on the cone so that
the Majorana fermions have a definite eigenvalue $h_s$ of 
spin helicity $({\bm k}_{\parallel}\times {\bm s})/k_{\parallel}$ at low energy.
%The spin-helicity is $h_s={\rm
%sgn}(v_{\rm surf})$ 
%[$h_s=-{\rm sgn}(v_{\rm surf})$] for the upper (lower) cone.  
A unique characteristic of Majorana fermions in $\Delta_2$ is that their spin helicity
varies depending on the chemical potential of the system.
As shown in Figs. \ref{Tdep}(a) and \ref{Tdep}(b), when the chemical potential $\mu_{\rm
STI}$ increases, the spin helicity of the upper cone near ${\bm
k}_{\parallel}={\bm 0}$ changes from
$h_s=-$ to $h_s=+$ at a critical value $\mu_{\rm STI}^{\rm c}$ \cite{yamakage12}. 
We also find that, when $\mu_{\rm STI}<\mu_{\rm
STI}^{\rm c}$, the energy
dispersion of the Majorana fermion is not a simple cone but a rather
complicated caldera-shaped one. 
As a result, in addition to the cone with $h_s=-$ near ${\bm
k}_{\parallel}={\bm 0}$, there appears another branch with $h_s=+$ in the
spectrum, as illustrated in Fig. \ref{Tdep}(b).
Referring to the spin helicity of the upper cone
near ${\bm k}_{\parallel}={\bm 0}$, we denote the STI with a simple cone 
and that with a complicated caldera-shaped one
as STI(+) and STI($-$), respectively.

Now, let us study the spin-helicity dependence of the Josephson current.
For this purpose, we consider the STI/STI junction illustrated in
Fig. \ref{Tdep}(c).
By tuning the chemical potentials in the 
left and right STIs, we can change the spin helicity of the Majorana cone
in each STI.
We calculate the Josephson current at STI/STI junctions with three
different combinations of spin helicity:
STI(+)/STI(+), STI(+)/STI($-$), and STI($-$)/STI($-$).
At these junctions, the current-phase relation of $J(\varphi)$ is rather conventional; however, the temperature dependence becomes anomalous.  
Figure \ref{Tdep}(d) shows  the temperature dependence of the 
maximum Josephson current, $\mathrm{max}_\varphi (J),$ for these junctions.
%%%%%%%%%%%%%%%%%%%%%%%%%%%%%%%%%%%%%%%%%%%%%%%%%%%%%%%%%%%%%
We find that, compared to junctions with the same spin helicity, 
 i.e., STI(+)/STI(+) and STI($-$)/STI($-$) junctions,
the Josephson current with the mismatched spin helicity
[STI(+)/STI($-$) junction] is strongly suppressed.
In particular, we find that 
the Josephson current at the latter junction decreases 
at low temperature whereas that at the former increases.
Here, we note that 
the suppression in the mismatched case occurs below $T/T_{\rm c STI}
 \sim 0.07$, which exactly corresponds to the energy $|E/\Delta|<0.07$ where 
the $h_s=-$ branch appears in STI($-$) [see Fig. \ref{Tdep}(b)].
This implies that the suppression of the Josephson current occurs owing to
 the mismatch of the spin helicity at the junction.  
%%%%%%%%%%%%%%%%
% \begin{figure}
% \centering
% \includegraphics{energy_STI_4.eps}
% \caption{Energy spectra of the surface states in STIs.
%  (a) 
% The positive energy part of the surface
%  state for $\mu_{\rm STI}=v_z$ in the region of $k_y=0$ and $k_x>0$.
% Inset shows the corresponding
%  overall spectrum of the $h_s=+$ branch of the surface state.
% (b)
% The positive energy part of the surface
%  state for $\mu_{\rm STI}=0.9v_z$ in the region of $k_y=0$ and $k_x>0$.
% Inset shows the corresponding overall spectrum of the $h_s=-$ branch of
%  the surface state, 
% which is discussed in Ref. \cite{yamakage12}.
% }
% \label{energy}
% \end{figure}
%%%%%%%%%%%%%%%
% \begin{figure}
% \centering
% \includegraphics[scale=0.25]{geometry_sti.eps}
% \caption{Geometry of STI/STI junction.}
% \label{geometry_sti}
% \end{figure}
%%%%%%%%%%%%%%%%
% \begin{comment}
We also find that 
%although the Josephson current with matched
%spin-helicities are enhanced at low temperature, 
the Josephson current at 
STI($-$)/STI($-$) is much more enhanced at low temperature than that at
STI(+)/STI(+) because a twisted energy spectrum of the caldera cone
has many low-lying states that contribute to the Josephson current.
%
%the magnitude of Josephson current becomes larger in the low
%temperature regime, as in Refs. \cite{asano03,asano11}, 
%
%On the other hand, in the C/C junction, 
%the enhancement of the magnitude of the Josephson current 
%at low temperature is weak. 
% since the 
%
%however, anomalous enhancement of Josephson current in the does not always occur%rs (See the case of C/C in Fig. \ref{Tdep}.) even if the superconductor support%s gapless surface states.
%
We mention here that the spin-helicity dependence of the
Josephson current is different from that of two-dimensional
helical superconductors because 
the Josephson current in two-dimensional helical superconductors is
always enhanced at low temperature, independent of the spin helicity
\cite{asano10}.

{\it Discussion.}
The anomalous Josephson effects in the topological phase 
reported in this Letter are accessible
experimentally. First, the vanishing of $\sin \varphi$ and $\cos
\varphi$ terms in the current-phase relation at $s/\Delta_2$ and
$d_{yz}/\Delta_2$ junctions is detectable by Shapiro steps at a
bias voltage of $V = n
\hbar \omega /(4e)$, where $n$ is an integer and $\omega$ is the microwave frequency.
DC SQUIDs with these junctions are also  sensitive to the
second-order behavior $\sin 2\varphi$ of the current-phase relation. 
Moreover, the anomalous temperature dependence of STI/STI junctions 
is easily measurable in experiments.
The anomalous temperature dependence is of particular interest because
it is a direct experimental
signal of the spin-locked nature of surface helical Majorana fermions.
The spin-helicity-dependent Josephson current is direct experimental evidence of the topological superconductivity of STIs.

This work is supported by
the ``Topological Quantum Phenomena" Grant-in Aid (No. 22103005) for Scientific Research on Innovative Areas from the Ministry of Education, Culture, Sports, Science and Technology (MEXT) of Japan.

\bibliography{josephson}
\clearpage
\end{document}